\documentclass[aps,pra,twocolumn,floatfix,superscriptaddress,amsmath,amssymb,eqsecnum]{revtex4}
\usepackage[breaklinks=true]{hyperref}
\usepackage{graphicx}
\usepackage{dsfont} 
\usepackage[caption=false]{subfig}

\newcommand{\Id}{\mathds{1}}

\DeclareMathOperator{\Real}{Re}
\DeclareMathOperator{\Imag}{Im}
\newcommand{\abs}[1]{\left\vert#1\right\vert}
\newcommand{\defined}{\mathrel{\mathop:}=}
\newcommand{\vect}[1]{\mathbf{#1}}
\newcommand{\evalat}[2]{\left.{#1}\right\vert_{#2}}
\DeclareMathOperator{\Tr}{Tr}
\newcommand{\ket}[1]{\left\vert{#1}\right\rangle}
\newcommand{\bra}[1]{\left\langle{#1}\right\vert}
\newcommand{\oprod}[2]{\left\vert{#1}\middle\rangle\middle\langle{#2}\right\vert}
\newcommand{\iprod}[2]{\left\langle{#1}\middle\vert{#2}\right\rangle}
\newcommand{\expt}[1]{\left\langle{#1}\right\rangle}

\newcommand{\frf}[1]{Fig.~\ref{#1}}
\newcommand{\Frf}[1]{Figure~\ref{#1}}
\newcommand{\srf}[1]{Sec.~\ref{#1}}

\hypersetup{
  colorlinks = true, 
  urlcolor   = blue, 
  linkcolor  = blue, 
  citecolor  = red 
}

\begin{document}

\title{On the novelty, efficacy, and significance of weak measurements for
  quantum tomography}

\author{Jonathan A.~Gross}\email{jagross@unm.edu}
\affiliation{Center for Quantum Information and Control, University of New
Mexico, Albuquerque NM 87131-0001, USA}
\author{Ninnat Dangniam}\email{ninnat@unm.edu}
\affiliation{Center for Quantum Information and Control, University of New
Mexico, Albuquerque NM 87131-0001, USA}
\author{Christopher Ferrie}\email{csferrie@gmail.com}
\affiliation{Centre for Engineered Quantum Systems, School of Physics, The University of Sydney, Sydney, NSW, Australia}
\author{Carlton M.~Caves}\email{ccaves@unm.edu}
\affiliation{Center for Quantum Information and Control, University of New
Mexico, Albuquerque NM 87131-0001, USA}
\affiliation{Centre for Engineered Quantum Systems, School of Mathematics and Physics,
University of Queensland, Brisbane, QLD 4072, Australia}
\date{\today}

\begin{abstract}

  The use of weak measurements for performing quantum tomography is
  enjoying increased attention due to several recent proposals. The advertised
  merits of using weak measurements in this context are varied, but are
  generally represented by novelty, increased efficacy, and foundational
  significance. We critically evaluate two proposals that make such claims and
  find that weak measurements are not an essential ingredient for most of their
  advertised features.
\end{abstract}

\maketitle

\section{Introduction}
\label{sec:intro}

The business of quantum state tomography is converting multiple copies of an
unknown quantum state into an estimate of that state by performing measurements
on the copies. The na\"ive approach to the problem involves measuring different
observables (represented by Hermitian operators) on each copy of the state and
constructing the estimate as a function of the measurement outcomes
(corresponding to different eigenvalues of the observables). Though tomography
can be performed in such a way, there are more general ways of interrogating the
ensemble; indeed, generalizations such as ancilla-coupled~\cite{Dar2002} and
joint~\cite{Mas1993} measurements lead one to evaluate the problem of tomography
from the perspective of \emph{generalized measurements}~\cite{Nie2010}, an
approach which has yielded many optimal tomographic
strategies~\cite{Hol1982,Hra1997,BagBalGil2006,Blu2010,Gro2010}.

An interesting subclass of generalized measurements is the class of \emph{weak
measurements\/}~\cite{BarLanPro1982, CavMil1987,DarYue1996, FucJac2001,
OreBru2005}. \Frf{fig:weak-measurement} gives a quantum-circuit description of a
weak measurement. Weak measurements are often the only means by which an
experimentalist can probe her system, thus making them of practical
interest~\cite{ChuGerKub1998,SmiSilDeu2006,GillettDaltonLanyon2010,SayDotZho2011,VijMacSli2012,CamFluRoc2013,CooRioDeu2014}.
Sequential weak measurements are also useful for describing continuous
measurements~\cite{WisemanMilburn2010}.

Weak measurements are also central in the more contentious formalism of
\emph{weak values\/}~\cite{AAV1988}. In particular, the technique of
\emph{weak-value amplification\/}~\cite{HostenKwiat2008} has generated much
debate over its metrological utility~\cite{StrubiBruder2013, Knee2013a,
TanYam2013, FerCom2013, Knee2013b, ComFerJia13a, ZhaDatWam13,
DresselMalikMiatto2014, JordanMartinez-RinconHowell2014, LeeTsutsui2014,
KneComFerGau14}.

The two proposals we review in this paper assert that it is useful to approach
the problem of tomography with weak measurements holding a prominent place in
one's thinking. Some care needs to be taken in identifying whether a
particular emphasis has the potential to be useful when thinking about
tomography, given the large body of work already devoted to the subject. Since
weak measurements are included in the framework of generalized measurements,
none of the known results for optimal measurements in particular scenarios are
going to be affected by shifting our focus to weak measurements. In
\srf{sec:eval-princ} we outline criteria for evaluating this shift of focus.

\begin{figure}[ht]
  \begin{center}
    \includegraphics[width=0.95\linewidth]{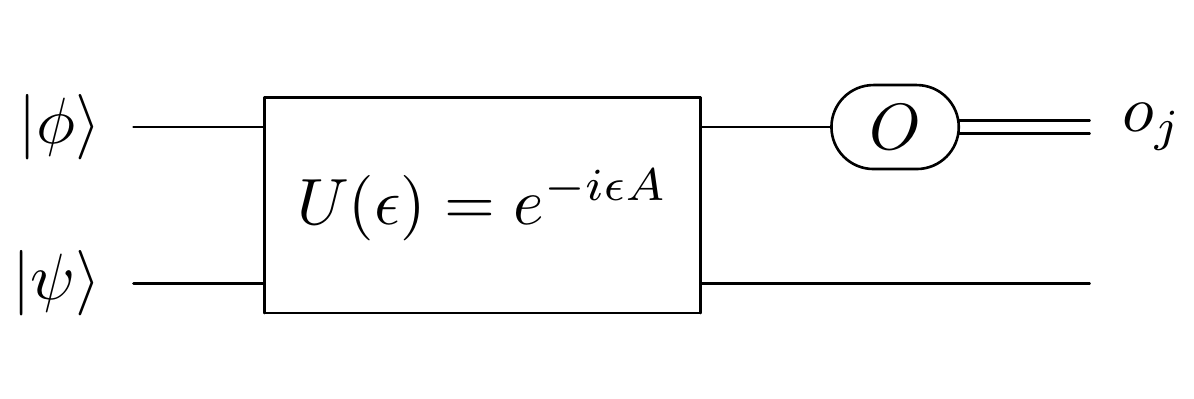}
  \end{center}
  \caption{A circuit depicting an ancilla-coupled measurement. Here
    $A$ is a two-system Hermitian operator, $\ket{\psi}$ is the state of the
    system being measured, $\ket{\phi}$ is the initial state of the meter,
    $\epsilon$ is a real number parameterizing the strength of the measurement, and $O$ is a standard observable with outcomes $o_j$.
    If $\abs{\epsilon}\ll1$ the measurement is weak, $U(\epsilon)\simeq\Id$,
    and very little is learned or disturbed about the system by measuring the
    meter.}
  \label{fig:weak-measurement}
\end{figure}

We apply these criteria to two specific tomographic schemes that advocate the
use of weak measurements.  \emph{Direct state tomography\/} (Sec.~\ref{sec:dst})
utilizes a procedure of weak measurement and postselection, motivated by
weak-value protocols, in an attempt to give an operational interpretation to
wavefunction amplitudes~\cite{LunSutPat2011}. \emph{Weak-measurement
tomography\/} (Sec.~\ref{sec:weak-tomo}) seeks to outperform so-called
``standard'' tomography by exploiting the small system disturbance caused by
weak measurements to recycle the system for further
measurement~\cite{DasArv2014}.

\section{Evaluation principles}
\label{sec:eval-princ}

Here we present our criteria for evaluating claims about the importance of weak
measurements for quantum state tomography.  The primary tool we utilize is
generalized measurement theory, specifically, describing a measurement by a
positive-operator-valued measure (POVM). A POVM assigns a positive operator
$E_F$ to every measurable subset $F$ of the set $\Omega$ of measurement outcomes
$\chi\in\Omega$.  For countable sets of outcomes, this means the measurement is
described by the countable set of positive operators,
\begin{align}
  \left\{ E_{\chi} \right\}_{\chi\in\Omega}\,.
  \label{eqn:countable-povm}
\end{align}
The positive operators $E_F$ are then given by the sums
\begin{align}
  E_F&=\sum_{\chi\in F}E_{\chi}\,.
  \label{eqn:povm-sums}
\end{align}
For continuous sets of outcomes the positive operator associated with a
particular measurable subset $F$ is given by the integral
\begin{align}
  E_F&=\int_FdE_\chi\,.
  \label{eqn:continuous-povm}
\end{align}
These positive operators capture all the statistical properties of a given
measurement, in that the probability of obtaining a measurement result $\chi$
within a measurable subset $F\subseteq\Omega$ for a particular state $\rho$ is
given by the formula
\begin{align}
  \Pr(\chi\in F|\rho)&=\Tr\big(\rho E_F \big)\,.
  \label{eqn:meas-stats}
\end{align}
That each measurement yields some result is equivalent to the completeness
condition,
\begin{align}
  E_\Omega&=\Id\,.
  \label{eqn:povm-completeness}
\end{align}

POVMs are ideal representations of tomographic measurements because they
contain all the information relevant for tomography, i.e., measurement
statistics, while removing many irrelevant implementation details. If two
wildly different measurement protocols reduce to the same POVM, their
tomographic performances are \hbox{identical}.

\subsection{Novelty}
\label{sec:eval-novel}

The authors of both schemes we evaluate make claims about the novelty of their
approach. These claims seem difficult to substantiate, since no tomographic
protocol within the framework of quantum theory falls outside the well-studied
set of tomographic protocols employing generalized measurements. To avoid
trivially dismissing claims in this way, however, we define a relatively
conservative subset of measurements that might be considered ``basic'' and ask
if the proposed schemes fall outside of this category.

The subset of measurements we choose is composed of randomly chosen
one-dimensional orthogonal projective measurements [hereafter referred to as
\emph{random ODOPs}; see \frf{fig:random-odop-limits}(a)]. These are the
measurements that can be performed using only traditional von Neumann
measurements, given that the experimenter is allowed to choose randomly the
observable he wants to measure.  This is quite a restriction on the full set of
measurements allowed by quantum mechanics.  Many interesting measurements, such
as symmetric informationally complete POVMs, like the tetrahedron measurement
shown in \frf{fig:random-odop-limits}(b), cannot be realized in such a way. With
ODOPs assumed as basic, however, if the POVM generated by a particular
weak-measurement scheme is a random ODOP, we conclude that weak measurements
should not be thought of as an essential ingredient for the scheme.

\begin{figure*}[t]
  \begin{center}
    \subfloat[]{
      \includegraphics[width=.6\textwidth]{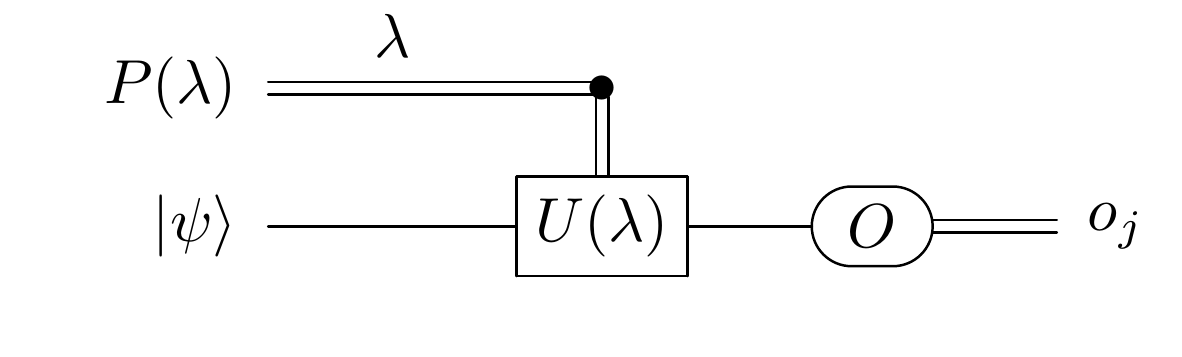}
    }
    \qquad
    \subfloat[]{
      \includegraphics[width=.3\textwidth]{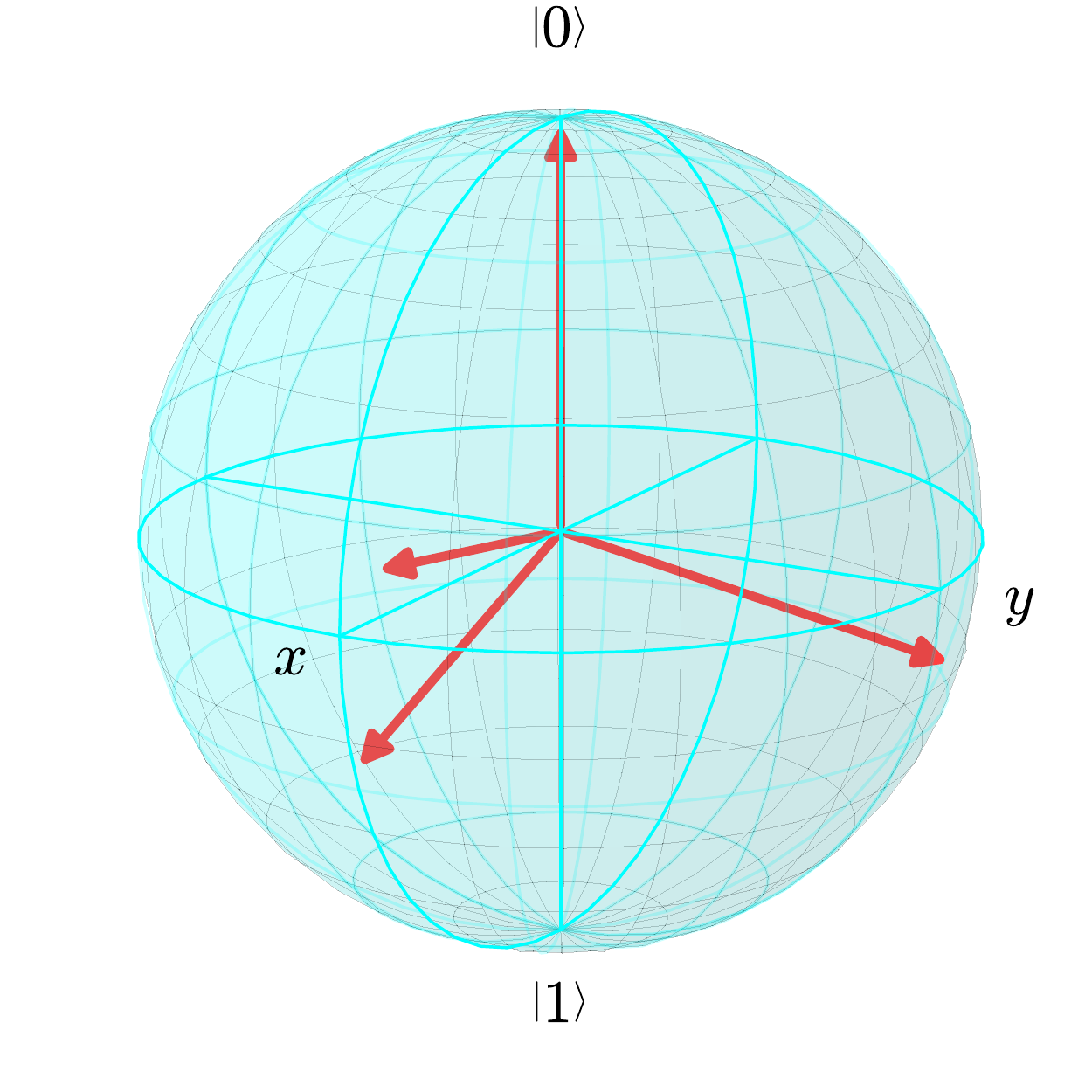}
    }
  \end{center}
  \caption{(a)~Implementation of a random ODOP by performing a randomly selected [probability $P(\lambda)$], basis-changing unitary $U(\lambda)$ before making a projective measurement of a standard observable $O$, with outcomes $o_j$.  (b)~POVM elements, represented as Bloch vectors, for the tetrahedron measurement, an example of a POVM that cannot be implemented as a random \hbox{ODOP}, because the POVM elements cannot be sorted into sets of equally weighted orthogonal projectors.}
  \label{fig:random-odop-limits}
\end{figure*}

Identifying other subsets of POVMs as ``basic'' might yield other interesting
lines of inquiry.  For example, when doing tomography on ensembles of atoms,
weak collective measurements might be compared with nonadaptive separable
projective measurements~\cite{SmiSilDeu2006,CooRioDeu2014}.

\subsection{Efficacy}

Users of tomographic schemes are arguably less interested in the novelty of a
particular approach than they are in its performance. There is a variety of
performance metrics available for state estimates, some of which have
operational interpretations relevant for particular applications. Given that we
have no particular application in mind, we adopt a reasonable figure of merit,
Haar-invariant average fidelity, which fortuitously is the figure of merit
already used to analyze the scheme we consider in \srf{sec:weak-tomo}. This is
the fidelity, $f\big(\rho,\hat\rho(\chi)\big)$, of the estimated state
$\hat{\rho}(\chi)$ with the true state $\rho$, averaged over possible
measurement records $\chi$ and further averaged over the unitarily invariant
(maximally uninformed) prior distribution over pure true states. For the case of
discrete measurement outcomes, this quantity is written as
\begin{align}
  F\big( \hat{\rho},E \big)
  &\defined\int d\rho\sum_\chi\Pr(\chi|\rho)
  f\big( \rho,\hat{\rho}(\chi) \big)\,.
  \label{eqn:avg-fidelity}
\end{align}

An obvious problem with this figure of merit is its dependence on the
estimator $\hat{\rho}$. We want to compare measurement schemes directly, not
measurement--estimator pairs. To remove this dependence we should calculate
the average fidelity with the optimal estimator for each measurement, expressed
as
\begin{align}
  F(E)&\defined\max_{\hat{\rho}}F\big(\hat{\rho},E\big)\,.
  \label{eqn:opt-est-avg-fidelity}
\end{align}

To avoid straw-man arguments, it is also important to compare the
performance of a particular tomographic protocol to the optimal protocol, or at
least the best known protocol. Both proposals we review in this paper are
nonadaptive measurements on each copy of the system individually. Since there
are practical reasons for restricting to this class of measurements, we
compare to the optimal protocol subject to this constraint.

This brings up an interesting point that can be made before looking at any of
the details of the weak-measurement proposals. For our chosen figure of merit,
the optimal individual nonadaptive measurement is a random ODOP (specifically
the Haar-invariant measurement, which samples a measurement basis from a uniform
distribution of bases according to the Haar measure). Therefore, weak-measurement
schemes cannot hope to do better than random ODOPs, and even if they are able
to attain optimal performance, weak measurements are clearly not an essential
ingredient for attaining that performance.

\subsection{Foundational significance}

Many proposals for weak-measurement tomography are motivated not by efficacy,
but rather by a desire to address some foundational aspect of quantum mechanics.
This desire offers an explanation for the attention these proposals receive in
spite of the disappointing performance we find when they are compared to random
ODOPs.

There are two prominent claims of foundational significance. The first is that a
measurement provides an operational interpretation of
wavefunction amplitudes more satisfying than traditional interpretations. This
is the motivation behind the direct state tomography of Sec.~\ref{sec:dst},
where the measurement allegedly yields expectation values directly proportional
to wavefunction amplitudes rather than their squares.

The second claim is that weak measurement finds a clever way to get around the
uncertainty--disturbance relations in quantum mechanics. The intuition behind
using weak measurements in this pursuit is that, since weak measurements
minimally disturb the system being probed, they might leave the system available
for further use; the information obtained from a subsequent measurement,
together with the information acquired from the preceding weak measurements,
might be more information in total than can be obtained with traditional
approaches.  Of course, generalized measurement theory sets limits on the amount
of information that can be extracted from a system, suggesting that such a
foundational claim is unfounded. We more fully evaluate this claim in
Sec.~\ref{sec:weak-tomo}.

\section{Direct state tomography}
\label{sec:dst}

In~\cite{LunSutPat2011} and~\cite{LunBab2012} Lundeen \emph{et al.}\ propose a
measurement technique designed to provide an operational interpretation of
wavefunction amplitudes. They make various claims about the measurement,
including its ability to make ``the real and imaginary components of the
wavefunction appear directly'' on their measurement device, the absence of a
requirement for global reconstruction since ``states can be determined locally,
point by point,'' and the potential to ``characterize quantum states \emph{in
situ\/} \ldots without disturbing the process in which they feature.''  The
protocol is thus often characterized as {\em direct state tomography\/} (DST).

To evaluate these claims, we apply the principles discussed in
Sec.~\ref{sec:eval-princ}. \ Lundeen \emph{et al.}\ have
outlined procedures for both pure and mixed states.  We focus on the pure-state
problem for simplicity, although much of what we identify is directly applicable
to mixed-state \hbox{DST}.  To construct the POVM, we need to describe the
measurement in detail. The original proposal for DST of Lundeen
\emph{et al.}\ calls for a continuous meter for performing the weak
measurements. As shown by Maccone and Rusconi~\cite{MacRus2014}, the continuous
meter can be replaced by a qubit meter prepared in the positive $\sigma_x$
eigenstate $\ket{+}$, a replacement we adopt to simplify the analysis.  Since
wavefunction amplitudes are basis-dependent quantities, it is necessary to
specify the basis in which we want to reconstruct the wavefunction. We call this
the \emph{reconstruction basis\/} and denote it by $\left\{ \ket{n}
\right\}_{0\leq n<d}$, where $d$ is the dimension of the system we are
reconstructing.

The meter is coupled to the system via one of a collection of interaction
unitaries $\left\{ U_{\varphi,n} \right\}_{0\leq n<d}$, where
\begin{align}
  U_{\varphi,n}&\defined e^{-i\varphi\oprod{n}{n}\otimes\sigma_z}\,.
  \label{eqn:dst-unitary}
\end{align}
The strength of the interaction is parametrized by $\varphi$.  A weak
interaction, i.e., one for which $|\varphi|\ll1$, followed by  measuring either
$\sigma_y$ or $\sigma_z$ on the meter, effects a weak measurement of the system.
In addition, after the interaction, there is a strong (projective) measurement
directly on the system in the \emph{conjugate basis\/}
$\left\{\ket{c_j}\right\}_{0\leq j<d}$, which is defined by
\begin{align}
  \iprod{n}{c_j}=\omega^{nj}/\sqrt d\;,\qquad
  \omega\defined e^{2\pi i/d}\,.
  \label{eqn:conjugate-basis}
\end{align}

The protocol for DST of Lundeen~\emph{et al.}, motivated by
thinking in terms of weak values, discards all the data except for the case when
the outcome of the projective measurement is $c_0$.  This protocol is depicted
as a quantum circuit in \frf{fig:dst-protocol}.

\begin{figure}[ht]
  \begin{center}
    \includegraphics[width=.9\linewidth]{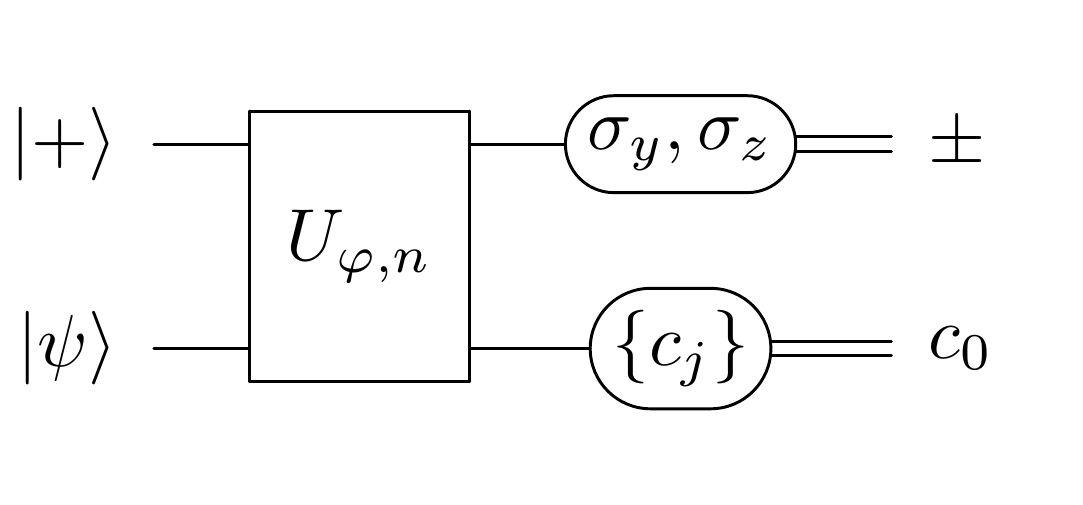}
  \end{center}
  \caption{Quantum circuit depicting direct state tomography. The meter is
    coupled to the system via one of a family of unitaries, $\left\{ U_{\varphi,n} \right\}_{0\leq n<d}$, each of which corresponds to a reconstruction-basis element.  The meter is then measured in either the $y$ or $z$ basis to obtain information about either the real or imaginary part
    of the wavefunction amplitude of the selected reconstruction-basis element.  This procedure is
    postselected on obtaining the $c_0$ outcome from the measurement of the
    system in the conjugate basis.  While the postselection is often described as
    producing an effect on the meter, the circuit makes clear that the
    measurements can be performed in either order, so it is equally valid to say
    the measurement of the meter produces an effect on the system.}
  \label{fig:dst-protocol}
\end{figure}

For each $n$, the expectation values of $\sigma_y$ and $\sigma_z$,
conditioned on obtaining the outcome $c_0$ from the projective
measurement, are given by
\begin{align}\label{eqn:sigmay}
\begin{split}
  \evalat{\expt{\sigma_y}}{n,c_0}
  &=\frac{2\sin\varphi}{d\,\Pr\!\big(c_0|U_{n,\varphi},\psi\big)}\Real(\psi_n\Upsilon^*)\\
  &\qquad+\frac{\sin2\varphi-2\sin\varphi}{d\,\Pr\!\big(c_0|U_{n,\varphi},\psi\big)}
  |\psi_n|^2\,,
\end{split}\\
  \evalat{\expt{\sigma_z}}{n,c_0}
  &=\frac{2\sin\varphi}{d\,\Pr\!\big(c_0|U_{n,\varphi},\psi\big)}
  \Imag(\psi_n\Upsilon^*)\,,
  \label{eqn:sigmaz}
\end{align}
where $\psi_n\defined\iprod{n}{\psi}$.  The probability for obtaining outcome
$c_0$ is
\begin{align}
\begin{split}
  &\Pr\!\big(c_0|U_{n,\varphi},\psi\big)\\
  &\;=\frac{1}{d}\Big(|\Upsilon|^2+
  2(\cos\varphi-1)\big[\Real(\psi_n\Upsilon^*)-|\psi_n|^2\big]\Big)\,,
\end{split}
\end{align}
and
\begin{align}\label{eqn:Upsilon}
\Upsilon&\defined\sum_n\psi_n\,.
\end{align}
We can always choose the unobservable global phase of $\ket\psi$ to make
$\Upsilon$ real and positive.  With this choice, which we adhere to going
forward, $\evalat{\expt{\sigma_y}}{n,c_0}$ provides information about the real
part of $\psi_n$, and $\evalat{\expt{\sigma_z}}{n,c_0}$ provides information
about the imaginary part of $\psi_n$.

Specializing these results to weak measurements gives
\begin{align}
  \psi_n
  &=\frac{\Upsilon}{2\varphi}\!\left(\evalat{\expt{\sigma_y}}{c_0,n}+
  i\!\evalat{\expt{\sigma_z}}{c_0,n} \right)+\mathcal{O}(\varphi^2)\,.
  \label{eqn:reconstruction-formula}
\end{align}
This is a remarkably simple formula for estimating the state $\ket{\psi}$! There
is, however, an important detail that should temper our enthusiasm. Contrary to
the claim in~\cite{LunBab2012}, this formula does not allow one to reconstruct
the wavefunction point-by-point (amplitude-by-amplitude in this case of a
finite-dimensional system), because one has no idea of the value of the
``normalization constant'' $\Upsilon$ until \emph{all\/} the wavefunction
amplitudes have been measured. This means that while ratios of wavefunction
amplitudes can be reconstructed point-by-point, reconstructing the amplitudes
themselves requires a global reconstruction.  Admittedly, this reconstruction is
simpler than commonly used linear-inversion techniques, but it comes at the
price of an inherent bias in the estimator, arising from the weak-measurement
approximation, as was discussed in~\cite{MacRus2014}.

The scheme as it currently stands relies heavily on postselection, a procedure
that often discards relevant data. To determine what information is being
discarded and whether it is useful, we consider the measurement statistics of
$\sigma_y$ and $\sigma_z$ conditioned on an arbitrary outcome $c_m$ of the
strong measurement.  To do that, we first introduce a unitary operator $Z$,
diagonal in the reconstruction basis, which cyclically permutes conjugate-basis
elements and puts phases on reconstruction-basis elements:
\begin{align}
  Z\ket{c_j}=\ket{c_{j+1}}\,,\qquad
  Z\ket{n}=\omega^n\ket{n}\,.
  \label{eqn:z-unitary}
\end{align}
As is illustrated in \frf{fig:postselect}, postselecting on outcome $c_m$ with
input state $\ket\psi$ is equivalent to postselecting on $c_0$ with input state
$Z^{-m}\ket\psi=\sum_n\omega^{-mn}\psi_n\ket n$.

\begin{figure}[h!]
  \begin{center}
    \subfloat[]{
      \includegraphics[width=.45\textwidth]{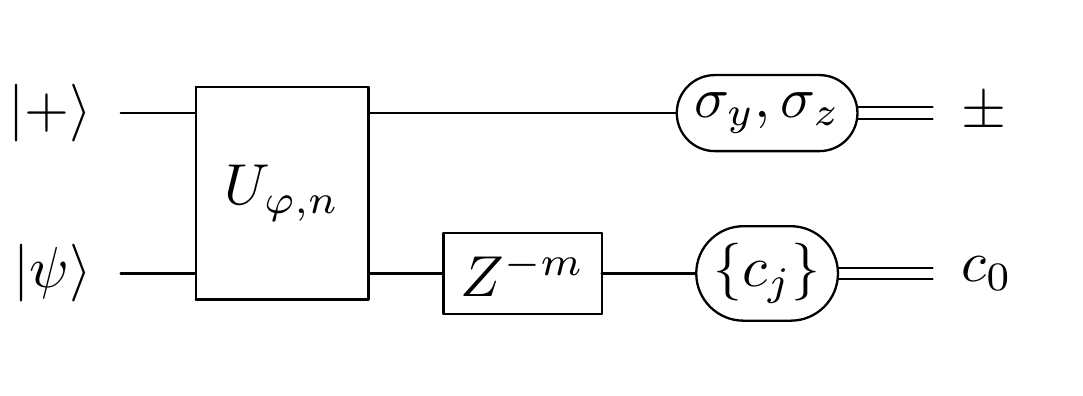}
    }
    \qquad
    \subfloat[]{
      \includegraphics[width=.45\textwidth]{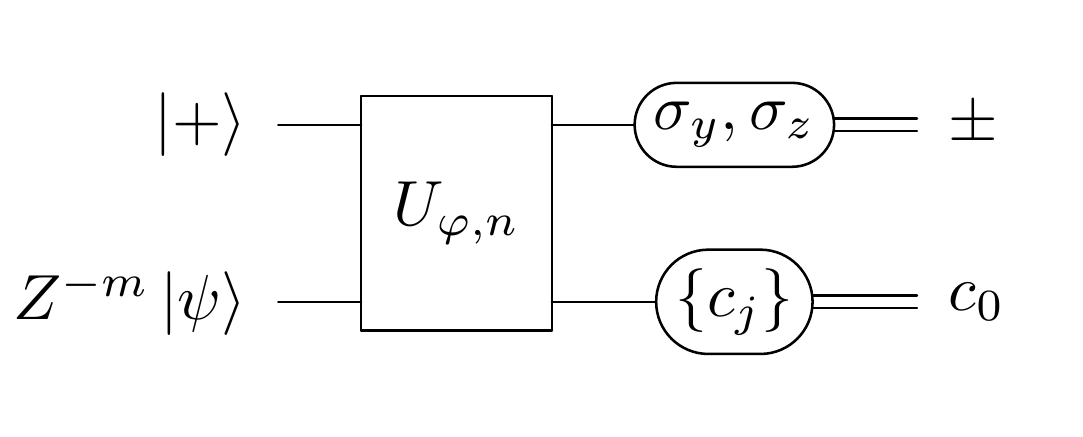}
    }
  \end{center}
  \caption{(a)~Postselection on outcome $c_m$, achieved by postselecting on
    $c_0$ after cyclic permutation of the conjugate basis by application of the
    unitary $Z^{-m}$.  (b)~Since $Z$ commutes with $U_{\varphi,n}$, (a)~is
    identical to postselecting on $c_0$ with input state $Z^{-m}\ket{\psi}$.}
  \label{fig:postselect}
\end{figure}

Armed with this realization, we can write reconstruction formulae for all
postselection outcomes,
\begin{align}
  \psi_n=\omega^{mn}\frac{\Upsilon}{2\varphi}\!
  \left( \evalat{\expt{\sigma_y}}{c_m,n}+
  i\!\evalat{\expt{\sigma_z}}{c_m,n} \right)+\mathcal{O}(\varphi^2)\,.
  \label{eqn:alt-reconstruction}
\end{align}
This makes it obvious that all the measurement outcomes in the conjugate basis
give ``direct'' readings of the wavefunction in the weak-measurement limit.
Postselection in this case is not only harmful to the performance of the
estimator, it is  not even necessary for the interpretational claims of \hbox{DST}.
Henceforth, we drop the postselection and include all the data produced by the
strong measurement.

The uselessness of postselection is not a byproduct of the use of a qubit meter.
In the continuous-meter case, the conditional expectation values in the
weak-measurement limit are given as weak values
\begin{align}
  \evalat{\expt{\oprod{x}{x}}}{p}&=\frac{\iprod{p}{x}\iprod{x}{\psi}}
  {\iprod{p}{\psi}}\,.
  \label{eqn:weak-value}
\end{align}
Weak-value-motivated DST postselects on meter outcome $p=0$ to hold the
amplitude $\iprod{p}{x}$ constant and thus make the expectation value
proportional to the wavefunction $\iprod{x}{\psi}$.  Since $\iprod{p}{x}$ is
only a phase, however, it is again obvious that postselecting on any value of
$p$ gives a ``direct'' reconstruction of a rephased wavefunction. Shi \emph{et
al.\/}~\cite{Shi2015} have developed a variation on Lundeen's protocol that
requires measuring weak values of only one meter observable.  This is made
possible by keeping data that is discarded in the original postselection
process.

We now consider whether the weak measurements in DST
contribute anything new to tomography.  It is already clear from
Eqs.~(\ref{eqn:sigmay}) and~(\ref{eqn:sigmaz}) that for this protocol to provide
data that is proportional to amplitudes in the reconstruction basis, the
weakness of the interaction is only important for the measurement of $\sigma_y$.
We are after something deeper than this, however, and to get at it, we change
perspective on the protocol of \frf{fig:dst-protocol}, asking not how
postselection on the result of the strong measurement affects the measurement of
$\sigma_y$ or $\sigma_z$, but rather how those measurements change the
description of the strong measurement.  As is discussed in
\frf{fig:dst-protocol}, this puts the protocol on a footing that resembles that
of the random ODOPs in \frf{fig:random-odop-limits}(a).

The measurement of $\sigma_z$, which provides the imaginary-part information, is
trivial to analyze, because the analysis can be reduced to drawing more
circuits.  In \frf{fig:imaginary-measurement}(a), the interaction unitary is
written in terms of system unitaries
$U_{n,\pm}\defined e^{\mp i\varphi\otimes\oprod{n}{n}}$ that are controlled in
the $z$-basis of the qubit.  The $\sigma_z$ measurement on the meter commutes
with the interaction unitary, so using the principle of deferred measurement, we
can move this measurement through the controls, which become classical controls
that use the results of the measurement.  The resulting circuit, depicted in
\frf{fig:imaginary-measurement}(b), shows that the imaginary part of each
wavefunction amplitude can be measured by adding a phase to that amplitude, with
the sign of the phase shift determined by a coin flip.  This is a particular
example of the random ODOP described by \frf{fig:random-odop-limits}(a).  We
conclude that weak measurements are not an essential ingredient for determining
the imaginary parts of the wavefunction amplitudes.

\begin{figure}[h!]
  \begin{center}
    \subfloat[]{
      \includegraphics[width=.4\textwidth]{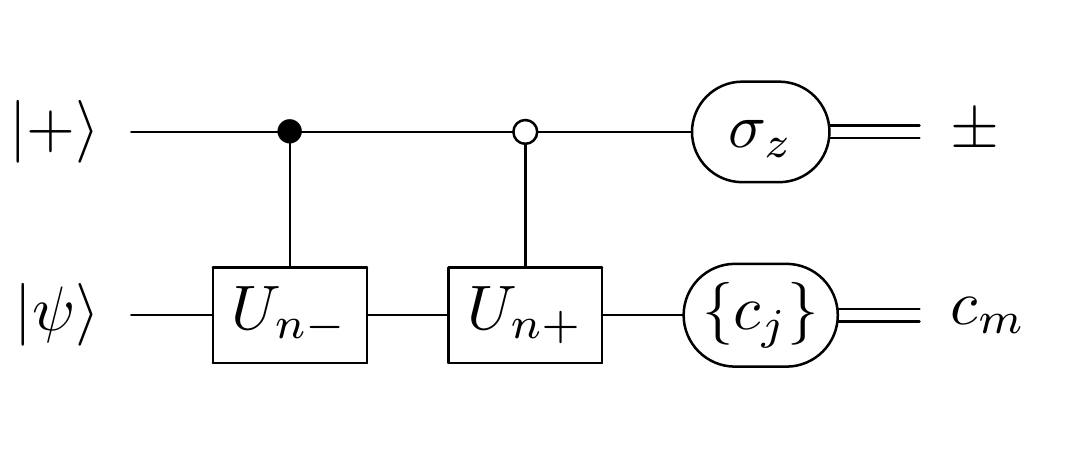}
    }
    \qquad
    \subfloat[]{
      \includegraphics[width=.5\textwidth]{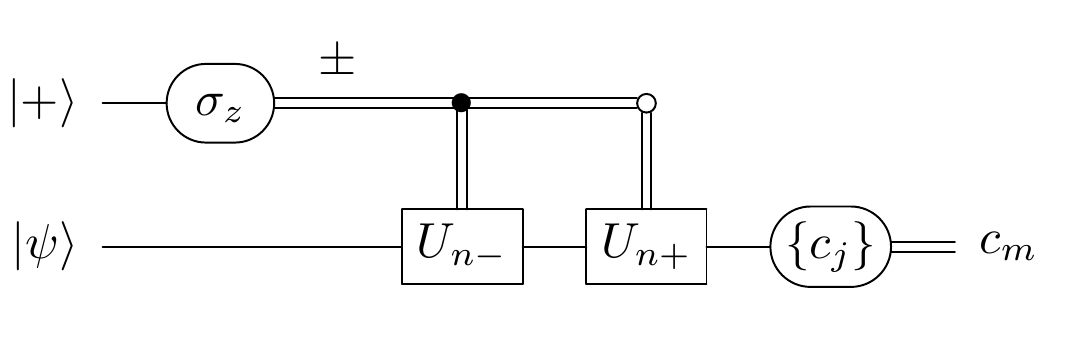}
    }
  \end{center}
  \caption{(a)~The imaginary-part measurement in a circuit where the interaction
    unitary is written in terms of system unitaries
    $U_{n,\pm}\defined e^{\mp i\varphi\oprod{n}{n}}$ controlled in the
    $z$-basis.  This circuit is equivalent to that of \frf{fig:dst-protocol}
    when $\sigma_z$ is measured on the meter; the result of the measurement
    reveals the sign of interaction, i.e., which of the unitaries $U_{n,\pm}$
    was applied to the system.  (b)~The coherent controls can be turned into
    classical controls by moving the measurement before the controls. The result
    is a particular instance of the random ODOP depicted in
    \frf{fig:random-odop-limits}(a).}	
  \label{fig:imaginary-measurement}
\end{figure}

Measuring the real parts is more interesting, since the $\sigma_y$ measurement
does not commute with the interaction unitary. We proceed by finding the Kraus
operators that describe the post-measurement state of the system.  The strong,
projective measurement in the conjugate basis has Kraus operators
$K_m=\oprod{c_m}{c_m}$, whereas the unitary interaction $U_{\varphi,n}$,
followed by the measurement of $\sigma_y$ with outcome $\pm$, has (Hermitian)
Kraus operator
\begin{align}
\begin{split}
  K_{\pm}^{(y,n)}&\defined\bra{\pm y}U_{\varphi,n}\ket{+}\\
  &=\frac{1}{\sqrt{2}}\left(\Id+\big(\sqrt2s_\pm-1\big)\oprod{n}{n}\right)\,,\\
  s_\pm&\defined\pm\sin(\varphi\pm\pi/4)\,,
\end{split}
\label{eqn:dst-kraus-ops}
\end{align}
where the eigenstates of $\sigma_y$ are
$\ket{\pm y}\defined\big(e^{\mp i\pi/4}\ket0+e^{\pm i\pi/4}\ket1\big)/\sqrt2$.
The composite Kraus operators, $K_{\pm,m}^{(y,n)}=K_mK_\pm^{(y,n)}$, yield POVM
elements $E_{\pm,m}^{(y,n)}\defined
K_{\pm,m}^{(y,n)\dagger}K_{\pm,m}^{(y,n)}=
K_\pm^{(y,n)}\oprod{c_m}{c_m}K_\pm^{(y,n)}$.
For each $n$, these POVM elements make up a rank-one POVM with $2d$ outcomes.

The POVM elements can productively be written as
\begin{align}\label{eqn:dst-povm}
  E_{\pm,m}^{(y,n)}&=\alpha^{(y)}_\pm\oprod{b_{\pm,m}^{(y,n)}}
  {b_{\pm,m}^{(y,n)}}\,,\\
  \begin{split}
  \ket{b_{\pm,m}^{(y,n)}}&\defined
  K_\pm^{(y,n)}\ket{c_m}\Big/\sqrt{\alpha^{(y)}_\pm}\\
  &=\left(\ket{c_m}+\big(\sqrt2 s_\pm-1\big)\omega^{mn}\ket{n}\right)\Big/
  \sqrt{2\alpha^{(y)}_\pm}\,,
  \end{split}\\
  \alpha^{(y)}_\pm&\defined\frac12\!\left(1-
  \frac{1}{d}+\frac{2}{d}s_\pm^2\right)\,.
\end{align}
The POVM for each $n$ does not fit into our framework of random ODOPs, but
can be thought of as within a wider framework of random POVMs.
Indeed, the Neumark extension~\cite{Neumark1940,Peres1993} teaches us how
to turn any rank-one POVM into an
ODOP in a higher-dimensional Hilbert space, where the dimension matches the
number of outcomes of the rank-one POVM.

Vallone and Dequal~\cite{ValDeq2015} have proposed an augmentation of the
original DST to obtain a ``direct'' wavefunction measurement without the need
for the weak-measurement approximation. The essence of their protocol is to
perform an additional $\sigma_x$ measurement on the meter. The statistics of
this measurement allow the second-order term in $\varphi$ to be eliminated from
the real-part calculation, giving a reconstruction formula that is
exact for all values of $\varphi$. Of course, the claim that the original DST
protocol ``directly'' measures the wavefunction is misleading, and directness
claims for Vallone and Dequal's modifications are necessarily more misleading.
Even ratios of real parts of wavefunction amplitudes no longer can be obtained
by ratios of simple expectation values, since these calculations rely on both
$\sigma_x$ \emph{and\/} $\sigma_y$ expectation values for different measurement
settings.

We analyze this additional meter measurement in the same way we analyzed the
$\sigma_y$ measurement. The Kraus operators corresponding to the meter
measurements are
\begin{align}
  K_{+}^{(x,n)}&\defined\bra{+}U_{\varphi,n}\ket{+}
  =\Id+\big(\cos\varphi-1\big)\oprod{n}{n}\,,\\
  K_{-}^{(x,n)}&\defined\bra{-}U_{\varphi,n}\ket{+}
  =\sin\varphi\oprod{n}{n}\,.
  \label{eqn:dst-x-kraus-ops}
\end{align}
The composite Kraus operators, $K_{\pm,m}^{(x,n)}=K_mK_\pm^{(x,n)}$, yield POVM
elements $E_{\pm,m}^{(x,n)}$ that can be written as
\begin{align}\label{eqn:dst-x-povm}
  E_{\pm,m}^{(x,n)}&=\alpha^{(x)}_\pm\oprod{b_{\pm,m}^{(x,n)}}{b_{\pm,m}^{(x,n)}}\,,\\
  \begin{split}
  \ket{b_{+,m}^{(x,n)}}
  &\defined K_+^{(x,n)}\ket{c_m}\Big/\sqrt{\alpha^{(x)}_+}\\
  &=\Big(\ket{c_m}+\big(\cos\varphi-1\big)\omega^{mn}\ket{n}\Big)\Big/\sqrt{\alpha^{(x)}_+}\,,
  \end{split}\\\label{eqn:bxnminusm}
  \ket{b_{-,m}^{(x,n)}}
  &\defined K_-^{(x,n)}\ket{c_m}\Big/\sqrt{\alpha^{(x)}_-}
  =\ket{n}\,,\\
  \alpha^{(x)}_+&\defined 1-\frac{\sin^2\!\varphi}{d}\,,\qquad
  \alpha^{(x)}_-\defined \frac{\sin^2\!\varphi}{d}\,.
\end{align}

It is useful to ponder the form of the POVM elements for the  $\sigma_y$ and
$\sigma_x$ measurements of the DST protocols.  For the original  DST protocol of
\frf{fig:dst-protocol}, without postselection, the only equatorial measurement
on the meter is of $\sigma_y$; the corresponding POVM elements, given by
Eq.~(\ref{eqn:dst-povm}), are nearly measurements in the conjugate basis, except
that the $n$-component of the conjugate basis vector is  changed in magnitude by
an amount that depends on the result of the  $\sigma_y$ measurement.  For the
augmented DST protocol of Vallone and Dequal, the additional POVM
elements~(\ref{eqn:dst-x-povm}), which come from the measurement of $\sigma_x$
on the meter, are quite different depending on the result of the $\sigma_x$
measurement.  For the result~$+$, the POVM element is similar to the POVM
elements for the measurement of $\sigma_y$, but with a different modification of
the $n$-component of the conjugate vector.  For the result~$-$, the POVM element
is simply a measurement in the reconstruction basis; as we see below, the
addition of the measurement in the reconstruction basis has a profound effect on
the performance of  the augmented DST protocol outside the region of weak
measurements, an effect unanticipated by the weak-value motivation.

Although claims regarding the efficacy of DST are rather nebulous, we consider
the negative impact of the weak-measurement limit on tomographic performance.
In doing so, we assume for simplicity that the system is a qubit, in some
unknown pure state that is specified by polar and azimuthal Bloch-sphere angles,
$\theta$ and $\phi$.  In this case we assume that the reconstruction basis is
the eigenbasis of $\sigma_z$; the conjugate basis is the eigenbasis
of~$\sigma_x$.

The method we use to evaluate the effect of variations in $\varphi$ is taken
from the work of de Burgh \emph{et al.}\ \cite{deBurgh2008}, which uses the
\emph{Cram\'er--Rao bound\/} (CRB) to establish an asymptotic (in number of
copies) form of the average fidelity.

\begin{figure}[h!]
  \begin{center}
    \includegraphics[width=1\linewidth]{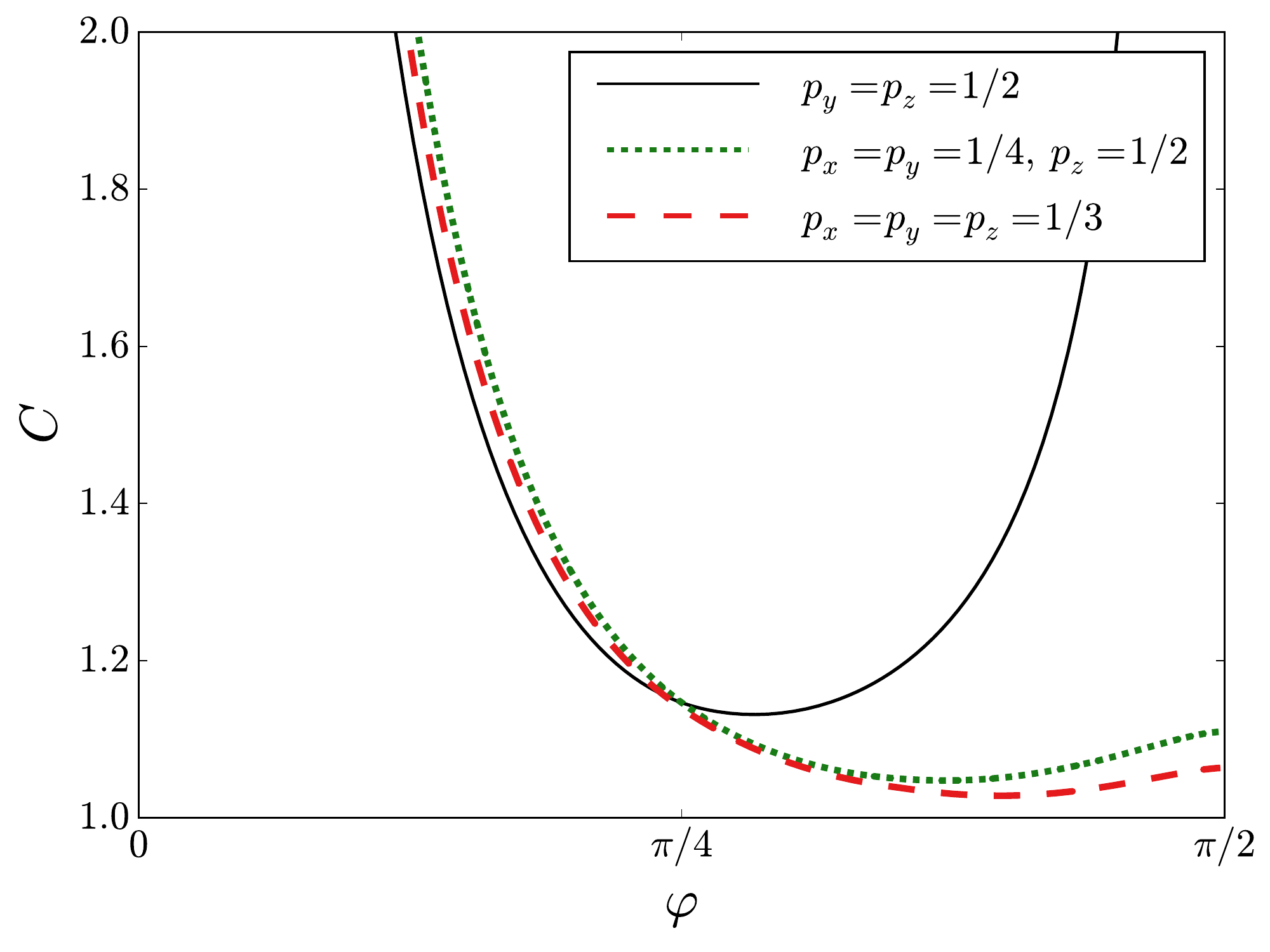}
  \end{center}
  \caption{(Color online) CRB $C(\varphi)$ of Eq.~(\ref{eqn:CRB}) for original
    DST (solid black) and augmented DST (dotted green for probability $1/4$ for
    $\sigma_x$ and $\sigma_y$ measurements and $1/2$ probability for a
    $\sigma_z$ measurement; dashed red for equal probabilities for all
    three measurements).  As the plot makes clear, the optimal values for
    $\varphi$ are far from the weak-measurement limit. Values of $\varphi$ for
    which $\varphi^2\simeq0$ give exceptionally large CRBs, confirming the
    intuition that weak measurements learn about the true state very slowly. The
    CRB for original DST also grows without bound as $\varphi$ approaches
    $\pi/2$, since that measurement strength leads to degenerate Kraus operators
    and a POVM that, not informationally complete, consists only of projectors
    onto $\sigma_x$ and $\sigma_y$ eigenstates of the system.  The CRB remains
    finite when the meter measurements are augmented with $\sigma_x$, since the
    resultant POVM at $\varphi=\pi/2$ then includes $\sigma_z$ projectors on the
    system [i.e., projectors in the reconstructions basis; see
    Eq.~(\ref{eqn:bxnminusm})], giving an informationally complete
    overall~\hbox{POVM}.}
  \label{fig:fisher-term}
\end{figure}

In analyzing the two DST protocols, original and augmented, we assume that the
two values of $n$ are chosen randomly with probability $1/2$.  For the original
protocol, we choose the $\sigma_z$ and $\sigma_y$ measurements with probability
$1/2$.  For the augmented protocol, we make one of two choices: equal
probabilities for the $\sigma_x$, $\sigma_y$, and $\sigma_z$ measurements or
probabilities of $1/2$ for the $\sigma_z$ measurement and $1/4$ for the
$\sigma_x$ and $\sigma_y$ measurements.  Formally, these assumed probabilities
scale the POVM elements when all of them are combined into a single overall \hbox{POVM}.

The asymptotic form involves the Fisher informations, $J_\theta$ and $J_\phi$,
for the two Bloch-sphere state parameters, calculated from the statistics of
whatever measurement we are making on the qubit.  The CRB already
assumes the use of an optimal estimator.  When the number of copies,
$N$, is large, the average fidelity takes the simple form
\begin{align}\label{eqn:asympt-avg-fid}
  F(\varphi)&\simeq 1-\frac{1}{N}C(\varphi)\,,\\
  \begin{split}
  C(\varphi)
  &=\int_{0}^{\pi}d\theta\,\sin\theta\\
  &\quad\times\int_{0}^{2\pi}d\phi\,\frac{1}{4}\!\!\left(
  \frac{1}{J_\theta(\theta,\phi,\varphi)}+
  \frac{\sin^2\theta}{J_\phi(\theta,\phi,\varphi)}\!\right)\,.
  \end{split}\label{eqn:CRB}
\end{align}
Though we have derived analytic expressions for the Fisher informations, it is
more illuminating to plot the CRB~$C(\varphi)$, obtained by numerical
integration (see \frf{fig:fisher-term}).  For original DST, the optimal value of
$\varphi$ is just beyond $\pi/4$, invalidating all qualities of ``directness''
that come from assuming $\varphi\ll1$.  For the augmented DST of Vallone and
Dequal, the optimal value of $\varphi$ moves toward $\pi/2$, even further
outside the region of weak measurements.  In both cases, $C(\varphi)$ blows up
at $\varphi=0$; for weak measurements, $C(\varphi)$ is so large that the
information gain is glacial.

We visualize this asymptotic behavior by estimating the average fidelity over
pure states as a function of $N$ using the sequential Monte Carlo technique \cite{smc_foot}, for various
protocols and values of $\varphi$.  \Frf{fig:optimal-dst-sim} plots these
results and shows how the average fidelity, for the optimal value of $\varphi$,
approaches the asymptotic form~(\ref{eqn:asympt-avg-fid}) as $N$ increases. We
note that the estimator used in these simulations is the estimator optimized for
average fidelity discussed in~\cite{BagBalGil2006}. If we were to use the
reconstruction formula proposed by Lundeen \emph{et al.}, the performance would
be worse.

\begin{figure}[ht]
  \begin{center}
    \includegraphics[width=1\linewidth]{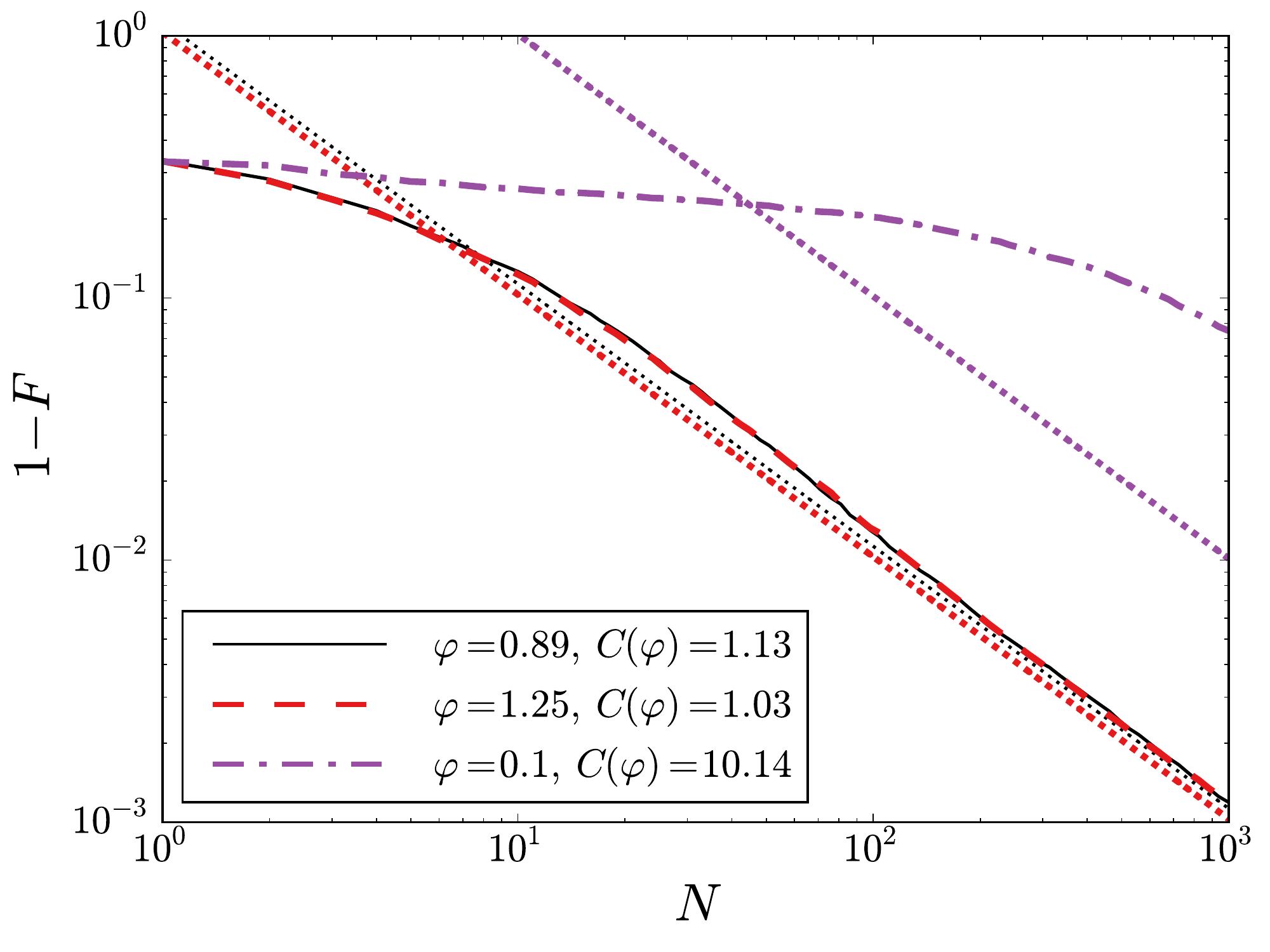}
  \end{center}
  \caption{(Color online) Average infidelity $1-F$ as a function of the number
  $N$ of system copies for three measurements.  The dashed red curve is
  for augmented DST with equal probabilities for the three meter measurements;
  the value $\varphi=1.25$ is close to the optimal value from
  \frf{fig:fisher-term}.  The other two curves are for original DST: the solid
  black curve is for $\varphi=0.89$, which is close to optimal (this curve
  nearly coincides with the dashed red curve for augmented DST); the
  dashed-dotted purple curve is for a small value $\varphi=0.1$, where the
  weak-measurement approximation is reasonable.  The three dotted curves give
  the corresponding asymptotic behavior $C(\varphi)/N$.  The two
  weak-measurement curves illustrate the glacial information acquisition when
  weak measurements are used; the dashed-dotted curve hasn't begun to approach
  the dotted asymptotic form for $N=10^3$.}
  \label{fig:optimal-dst-sim}
\end{figure}

Our conclusions are the following.  First, postselection contributes nothing to
\hbox{DST}.  Its use comes from attention to weak values, but postselection is
actually a negative for tomography because it discards data that are just as
cogent as the data that are retained in the weak-value scenario.  Second, weak
measurements in this context add very little to a tomographic framework based on
random ODOPs.  Finally, the ``direct'' in DST is a misnomer~\cite{misnomer}
because the protocol does not provide point-by-point reconstruction of the
wavefunction.

The inability to provide point-by-point reconstruction is a symptom of a general
difficulty.  Any procedure, classical or quantum, for detecting a complex
amplitude when only absolute squares of amplitudes are measurable involves
interference between two amplitudes, say, $A$ and $B$, so that some observed
quantity involves a product of two amplitudes, say, ${\rm Re}(A^*B)$.  If one
regards $A$ as ``known'' and chooses it to be real, then ${\rm Re}(B)$ can be
said to be observed directly.  This is the way amplitudes and phases of
classical fields are determined using interferometry and square-law detectors.

Of course, quantum amplitudes are not classical fields.  One loses the ability to
say that one amplitude is known and objective, with the other to be determined
relative to the known amplitude.  Indeed, if one starts from the tomographic
premise that nothing is known and everything is to be estimated from measurement
statistics, then $A$ cannot be regarded as ``known.''  DST fits into this
description, with the sum of amplitudes, $\Upsilon$ of Eq.~(\ref{eqn:Upsilon}),
made real by convention, playing the role of $A$.   The achievement of DST is
that this single quantity is the only ``known'' quantity needed to construct all
the amplitudes $\psi_n$ from measurement statistics.  Single quantity or not,
however, $\Upsilon$ must be determined from the entire tomographic procedure
before any of the amplitudes $\psi_n$ can be estimated.

\section{Weak-measurement tomography}
\label{sec:weak-tomo}

The second scheme we consider is a proposal for qubit tomography by Das and
Arvind~\cite{DasArv2014}.   This protocol was advertised as opening up ``new
ways of extracting information from quantum ensembles'' and outperforming, in
terms of fidelity, tomography performed using projective measurements on the
system.  The optimality claim cannot be true, of course, since a random ODOP
based on the Haar invariant measure for selecting the ODOP basis is optimal when
average fidelity is the figure of merit, but the novelty of the information extraction
remains to be evaluated.

\begin{figure}[ht!]
  \begin{center}
    \includegraphics[width=.95\linewidth]{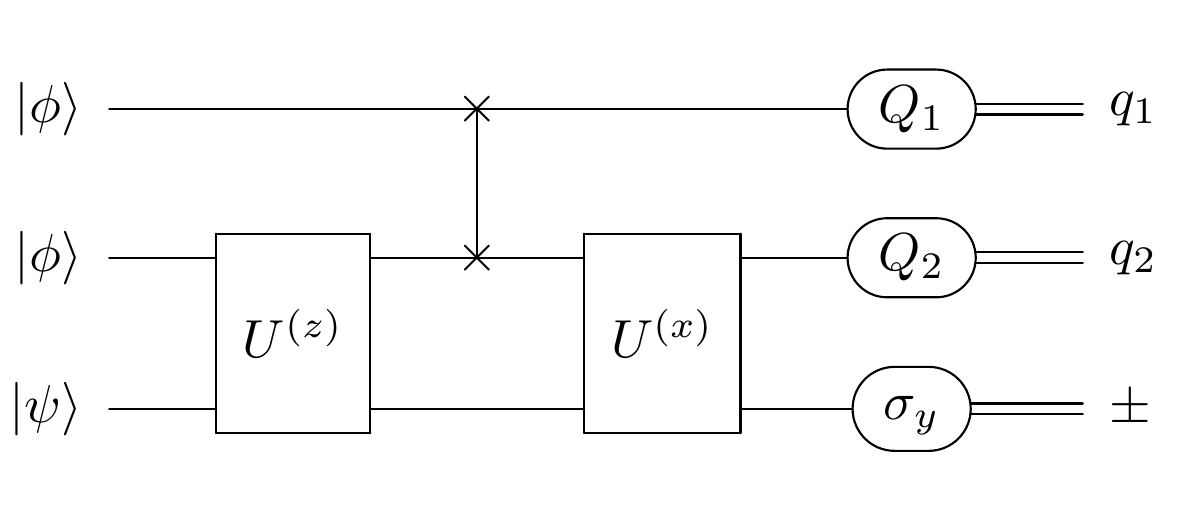}
  \end{center}
  \caption{Quantum circuit depicting the weak-measurement tomography protocol
    of Das and Arvind. Two identical meters are used as ancillas to perform the
    weak $z$ and $x$ measurements. The circuit makes clear that there is
    nothing important in the order the measurements are performed after the
    interactions have taken place, so we consider the protocol as a single
    ancilla-coupled measurement.}
  \label{fig:weak-protocol}
\end{figure}

The weak measurements in this proposal are measurements of Pauli components of
the qubit.  These measurements are performed by coupling the qubit system via an
interaction unitary,
\begin{align}
  U^{(j)}=e^{-i\sigma_j\otimes P}\,,
  \label{eqn:weak-meas-unitary}
\end{align}
to a continuous meter, which has position $Q$ and momentum $P$ and is prepared
in the Gaussian state
\begin{align}
  \ket{\phi}=\sqrt[4]{\frac{\epsilon}{2\pi}}\int_{-\infty}^{\infty}dq\,e^{-\epsilon q^2/4}\ket{q}\,.
  \label{eqn:weak-meas-Gaussian}
\end{align}
The position of the meter is measured to complete the weak measurement. The
weakness of the measurement is parametrized by $\epsilon=1/\Delta q^2$.

\begin{figure*}[t]
  \begin{center}
    \includegraphics[width=0.9\textwidth]{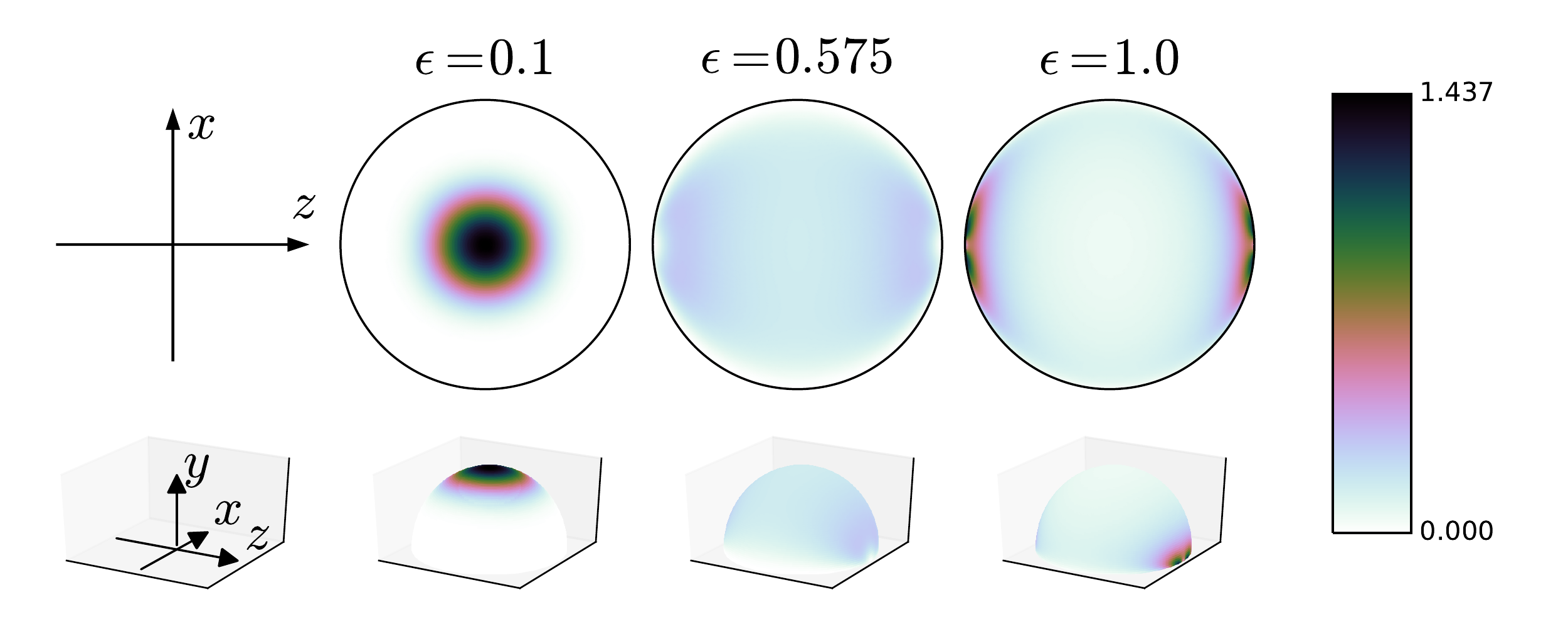}
  \end{center}
  \caption{(Color online) Effective distributions over measurement bases
    visualized as distributions over the positive-$y$ Bloch hemisphere. Very
    weak measurements of $z$ and $x$ (e.g., $\epsilon=0.1$) don't perturb the
    final $y$ measurement very much, so the distribution of bases is
    concentrated about the $y$ axis.  Very strong measurements (e.g.,
    $\epsilon=1$) end up extracting most of the information in the first $z$
    measurement, so the distribution becomes concentrated around the $z$ axis.
    The optimal measurement ($\epsilon\simeq0.575$) has an effective
    distribution that is nearly uniform.}
  \label{fig:bloch-plots}
\end{figure*}

The Das-Arvind protocol involves weakly measuring the $z$ and $x$ Pauli
components and then performing a projective measurement of $\sigma_y$.  We
depict this protocol as a circuit in \frf{fig:weak-protocol}.  Das and Arvind
view this protocol as providing more information than is available from the
projective $\sigma_y$ measurement because the weak measurements extract a little
extra information about the $z$ and $x$ Pauli components without appreciably
disturbing the state of the system before it is slammed by the projective
$\sigma_y$ measurement.  Again, we turn the tables on this point of view, with
its notion of a little information flowing out to the two meters, to a
perspective akin to that of the random ODOP of \frf{fig:random-odop-limits}(a).
We ask how the weak measurements modify the description of the final projective
measurement.  For this purpose, we again need Kraus operators to calculate the
POVM of the overall \hbox{measurement}.

The Kraus operators for the projective measurement are $K^{(y)}_\pm=\oprod{\pm
y}{\pm y}$, and the (Hermitian) Kraus operator for a weak measurement with
outcome $q$ on the meter~is
\begin{align}
\begin{split}
  K^{(j)}(q)&\defined\bra{q}\,U^{(j)}\ket{\phi}\sqrt{dq}\\
  &=\sqrt[4]{\frac{\epsilon}{2\pi}}\exp\!\left(-\frac{\epsilon(q^2+1)}{4}\right)\\
  &\quad\times\Big(\Id\cosh(\epsilon q/2)+\sigma_j\sinh(\epsilon q/2)\Big)
  \sqrt{dq}\,.
\end{split}
\label{eqn:weak-kraus-ops}
\end{align}
The Kraus operators for the whole measurement procedure are
$K_\pm(q_1,q_2)\defined K^{(y)}_\pm K^{(x)}(q_2)K^{(z)}(q_1)$. From these come
the infinitesimal POVM elements for outcomes $q_1$, $q_2$, and $\pm$:
\begin{align}
\begin{split}
  d&E_\pm(q_1,q_2)\\
  &\,\defined K^\dagger_\pm(q_1,q_2)K_\pm(q_1,q_2)\\
  &\,=K^{(z)}(q_1)K^{(x)}(q_2)\oprod{\pm y}{\pm y}K^{(x)}(q_2)K^{(z)}(q_1)\,.
\end{split}
\end{align}
These POVM elements are clearly rank-one.

Using the Pauli algebra, we can bring the POVM elements into the explicit form,
\begin{align}
\begin{split}
  dE_\pm(q_1,q_2)&=K^\dagger_\pm(q_1,q_2)K_\pm(q_1,q_2)\\
  &=dq_1\,dq_2\,G(q_1,q_2)
  \frac{1}{2}(\Id+\hat{\vect{n}}_\pm(q_1,q_2)\cdot\boldsymbol{\sigma})\,,
\end{split}
\label{eqn:weak-continuous-povm}
\end{align}
where we have introduced a probability density and unit vectors,
\begin{align} \label{eqn:prob-dens}
\begin{split}
  G(q_1,q_2)&\defined
  \frac{\epsilon}{2\pi}\exp\!\left(-\frac{\epsilon(q_1^2+q_2^2+2)}{2}\right)\\
  &\qquad\times\cosh\epsilon q_1\cosh\epsilon q_2\,,
\end{split}\\
  \hat{\vect{n}}_\pm(q_1,q_2)&\defined
  \frac{\hat{\vect x}\sinh\epsilon q_2
  \pm\hat{\vect y}
  +\hat{\vect z}\sinh\epsilon q_1\cosh\epsilon q_2}
  {\cosh\epsilon q_1\cosh\epsilon q_2}\,.
  \label{eqn:unit-vec}
\end{align}
We note that $G(q_1,q_2)=G(-q_1,-q_2)$ and
$\hat{\vect{n}}_\pm(q_1,q_2)=-\hat{\vect{n}}_\mp(-q_1,-q_2)$.  This means that
the overall POVM is made up of a convex combination of equally weighted pairs of
orthogonal projectors and is therefore a random \hbox{ODOP}.  From this perspective,
the weak measurements are a mechanism for generating a particular distribution
from which different projective measurements are sampled; i.e., they are a
particular way of generating a distribution $P(\lambda)$ in
\frf{fig:random-odop-limits}.  Several of these distributions are plotted in
\frf{fig:bloch-plots}.

\begin{figure}[ht!]
  \begin{center}
    \includegraphics[width=1\linewidth]{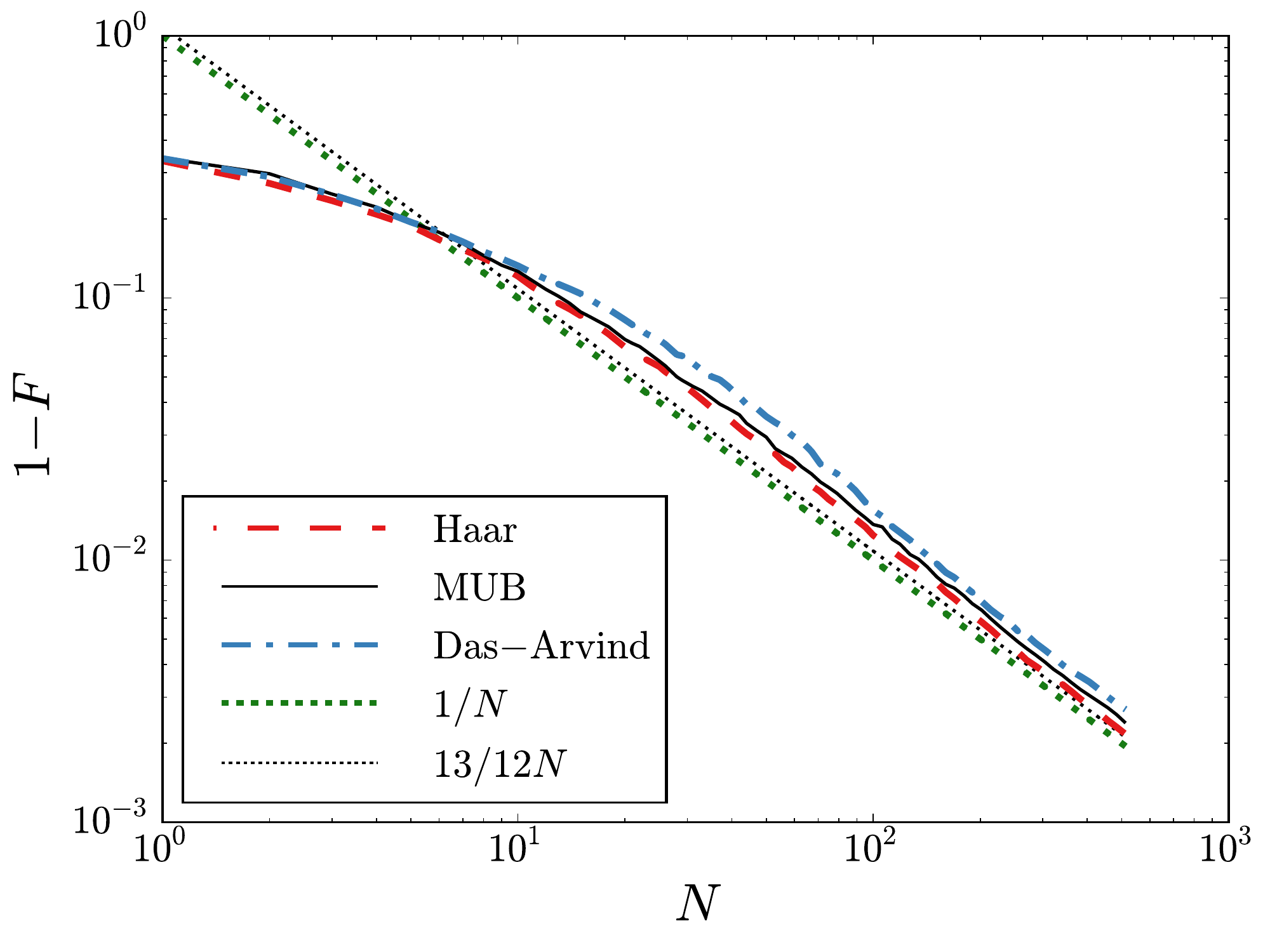}
  \end{center}
  \caption{(Color online) Average infidelity $1-F$ as a function of the number
  $N$ of system copies for three measurements: Das and Arvind's measurement
  protocol (dotted-dashed blue) with $\epsilon=0.575$; MUB consisting of Pauli
  $\sigma_x$, $\sigma_y$, and $\sigma_z$ measurements (solid black); random ODOP
  consisting of projective measurements sampled from the Haar-uniform
  distribution (dashed red).  The dotted lines, $1/N$ and $13/12N$, are the CRBs
  defined by Eq.~(\ref{eqn:CRB}) for the optimal generalized tomographic
  protocol and MUB measurements, respectively.}
  \label{fig:weak-mub-haar-sim}
\end{figure}

It is interesting to note that the value of $\epsilon$ that Das and Arvind
identified as optimal (about $0.575$) produces a distribution that is nearly
uniform over the Bloch sphere. This matches our intuition when thinking of the
measurement as a random ODOP, since the optimal random ODOP samples from the
uniform distribution

To visualize the performance of this protocol, we again use sequential
Monte Carlo simulations of the average fidelity. Das and Arvind compare their
protocol to a measurement of $\sigma_x$, $\sigma_y$, and $\sigma_z$, whose
eigenstates are \emph{mutually unbiased bases\/} (MUB).  In
\frf{fig:weak-mub-haar-sim}, we compare Das and Arvind's protocol for
$\epsilon=0.575$ to a MUB measurement and to the optimal
projective-measurement-based tomography scheme, i.e., the Haar-uniform
random~\hbox{ODOP}.

We don't discuss the process of binning the position-measurement results
that Das and Arvind engage in, since such a process produces a rank-2 POVM that
is equivalent to sampling from a discrete distribution over projective
measurements and then adding noise, a practice that necessarily degrades
tomographic performance.

We conclude that the protocol does not offer anything beyond that offered by
random ODOPs and that its claim of extracting information about the system
without disturbance is not supported by our analysis.  In particular, when
operated optimally, it is essentially the same as the strong projective measurements
of a Haar-uniform random \hbox{ODOP}. It is true that the presence of the
$z$ and $x$ measurements provides more information than a projective $y$
measurement by itself, but this is not because the $z$ and $x$ measurements
extract information without disturbing the system.

\section{Summary and conclusion}

Our analysis of weak-measurement tomographic schemes gives us guidance for
future forays into tomography.

POVMs contain the necessary and sufficient information for comparing the
performance of tomographic techniques. Specific realizations of a POVM might
provide pleasing narratives, but these narratives are irrelevant for calculating
figures of merit.  Optimal POVMs for many figures of merit and technical
limitations are known. A new tomographic proposal should identify restrictions
on the set of available POVMs that come about from practical considerations and
compare itself to the best known POVM in that set. The question of the
optimality of Das and Arvind's tomographic scheme is easily answered by
identifying what POVMs arise from ``projective measurement-based tomography''
and realizing these POVMs are optimal even in the generalized nonadaptive,
individual-measurement scenario.

Claims about novel properties of a state-reconstruction technique should be
evaluated as a comparison with a motivated restriction on the set of POVMs. The
false dichotomy between ``tomographic methods'' and whatever new method is being
proposed obfuscates that all new methods implement a POVM and that
reconstructing a state from POVM statistics is nothing but tomography. Our
analysis shows that even the relatively bland and conceptually simple set of
random ODOPs captures most of the behavior exhibited by more exotic
protocols.

It is appropriate to move beyond the minimal, platform-independent POVM
description when considering ease of implementation or when trying to provide a
helpful conceptual framework.  Nonetheless, a pleasing conceptual framework
should not be confused with an optimal experimental arrangement. If the
experimental setup described by Lundeen \emph{et al.}\ happens to be the easiest
to implement in one's lab, the state should still be reconstructed using
techniques developed in the general POVM setting rather than the perturbative
reconstruction formula presented in work on DST, regardless of how attractive
one finds the wavefunction-amplitude analogy.

\acknowledgements 
We thank Josh Combes for helpful discussions.  This work was support by National Science Foundation Grant No.~PHY-1212445.  CF was also supported by the Canadian Government through the NSERC PDF program, the IARPA MQCO program, the ARC via EQuS Project No.~CE11001013, and by the U.S.\ Army Research Office Grant Nos.~W911NF-14-1-0098 and~W911NF-14-1-0103.


\begin{thebibliography}{99}
\bibitem{misnomer}
In the long tradition of institutions abandoning a name in favor of initials, to
divorce from some original product or purpose, we recommend the use of DST in
the hope that the ``direct'' can be forgotten.
\end{thebibliography}
\end{document}